\documentclass[]{jfm}

\usepackage{graphicx}
\usepackage{newtxtext}
\usepackage{newtxmath}
\usepackage{natbib}
\usepackage{hyperref}
\hypersetup{
    colorlinks = true,
    urlcolor   = blue,
    citecolor  = black,
}

\newcommand{\RomanNumeralCaps}[1]
\linenumbers

\newcommand{\hu}{\hat{u}}
\newcommand{\hv}{\hat{v}}
\newcommand{\hw}{\hat{w}}
\newcommand{\hp}{\hat{p}}
\newcommand{\htheta}{\hat{\theta}}

\title{Non-Hermitian topological wall modes in rotating Rayleigh--B\'enard convection}

\author{Furu Zhang\aff{1}
 \and Jin-Han Xie\aff{1,2}
   \corresp{\email{jinhanxie@pku.edu.cn}}}

\affiliation{\aff{1} Department of Mechanics and Engineering Science at College of Engineering and State Key Laboratory for Turbulence and Complex Systems, Peking University, Beijing, China
\aff{2} Joint Laboratory of Marine Hydrodynamics and Ocean Engineering, Laoshan Laboratory, Shandong, China}

\begin{document}
\maketitle

\begin{abstract}
We show that the rotating Rayleigh–B\'enard convection, where a rotating fluid is heated from below, exhibits non-Hermitian topological states.
Recently, \cite{Favier_Knobloch.2020} hypothesized that the robust wall modes in rapidly rotating convection are topologically protected.
We study the linear problem around the conduction profile, and by considering a Berry curvature defined in the complex wavenumber space, particularly, by introducing a complex vertical wavenumber, we find that these modes can be characterized by a non-zero integer Chern number, indicating their topological nature.
The eigenvalue problem is intrinsically non-Hermitian, therefore the definition of Berry curvature generalizes that of the stably stratified problem.
Moreover, the three-dimensional setup naturally regularizes the eigenvector at the infinite horizontal wavenumber.
Under the hydrostatic approximation, it recovers a two-dimensional analogue of the one which explains the topological origin of the equatorial Kelvin and Yanai waves \citep{Delplace2017}.
The existence of the tenacious wall modes relies only on rotation when the fluid is stratified, no matter whether it is stable or unstable.
However, the neutrally stratified system does not support a topological edge state.
In addition, we define a winding number to visualize the topological nature of the fluid.

\end{abstract}

\begin{keywords}
Convection, Rotating flows, Topological fluid dynamics
\end{keywords}

\section{Introduction}

\label{sec:Introduction}

Rotating Rayleigh-B\'enard convection, where rotating fluid between two horizontal plates is driven by bottom heating, is an important prototype for geophysics and astrophysics \citep{annurev.fluid,RevModPhys.2009, king_stellmach_aurnou_2012, guervilly_hughes_jones_2014}. 
It occurs in various natural systems, such as the Earth's atmosphere and oceans, as well as in the interiors of stars and exoplanets.

In this convection system, the rotation leads to the emergence of edge states at the sidewall \citep{PhysRevLett.1991Asymmetric,zhong_ecke_steinberg_1993,Ecke_1992,goldstein_knobloch_mercader_net_1993,PhysRevE.47.3326}, which are characterized by their unidirectional propagation and their ability to persist even in the presence of different types of barriers and turbulence \citep{Favier_Knobloch.2020,ECKE2023Bits,Ecke2023Turbulent}. 
These states were first discovered through experiments indirectly by \cite{rossby_1969}, where a surprising convection occurs with a  Rayleigh number ($R_{a}$) below the critical value for the laterally unbounded system \citep{stability.1961}.
\cite{Zhang_Liao.2009} gives the asymptotic expression of the critical $R_{a}$ for the onset of these wall modes, regaining the leading terms previously calculated by \cite{herrmann_busse_1993}.
With increasing $R_{a}$, more wall modes emerge, and they may undergo modulational instabilities and interact with bulk modes, resulting in complex nonlinear dynamics \citep{zhong_ecke_steinberg_1993,Liu.prE.1999,horn_schmid_2017}.
In the turbulent flows with high $R_{a}$, the wall modes may be transformed to the boundary zonal flows, which can also coexist with bulk convection \citep{PhysRevLett.2020,PhysRevFluids.2020,PhysRevFluids.2022}.
Beyond these detailed studies, a new insight is proposed that these edge states may be related to the topological nature of the system \citep{Favier_Knobloch.2020}.
Thus, a new set of questions arises, such as, can we obtain the topological invariant of the system? How robust in quantitative terms is the topology in the presence of turbulence? 
The answers to these follow-up questions will deepen our understanding of the overall flow structure and the spatial distribution of heat flux in geophysical and astrophysical contexts \citep{PRL.2009.Zhong,PhysRevLett.2023}.

Edge states under topological protection have been a topic of great interest in recent years. These edge states are robust against disorder and perturbations, and their topological protection originates from the nontrivial topology of the underlying bulk system. 
This topological nature can be indicated by a corresponding topological invariant, such as the Chern number \citep{xiao2010,Delplace2017} or the winding number \citep{winding_number2023}. When the topological invariant is non-zero, the bulk-boundary correspondence guarantees the existence of robust edge states \citep{PRB2011,raytracing2023}.
The topological physics first appeared in condensed matters \citep{kosterlitz1973ordering,thouless1982quantized,Haldane1988}, such as topological insulators \citep{TIs2010}, topological photonics \citep{Tphoto2019} and topological phononics \citep{Tphono2020}. 
Recently similar topological properties have been found in macroscopic systems, especially in hydrodynamics, such as the topological origin of trapped waves near the equator or coastlines \citep{Delplace2017,PRR2021}, the topological waves in fluids with odd viscosity \citep{odd2019prl}, and the presence of topological invariants in active matter systems \citep{shankar2022nature}.
This universality of topological states can be attributed to the similar symmetries embodied in the systems of different scales \citep{senthil2015symmetry}.

In this paper, we point out that the linearized rotating convection is a non-Hermitian system and that these edge states are manifestations of the nontrivial topological Berry phase in the bulk. 
Non-Hermitian systems are open systems that do not obey the Hermitian symmetry property \citep{Non-Hermitian.physycs2020}. 
Thus, their eigenvalues are not necessarily real, and the corresponding eigenvectors can be unorthogonal. 
Rayleigh-Bénard convection is non-Hermitian because it exchanges energy with an external heat source.
Non-Hermitian systems exhibit unconventional physical properties, including non-reciprocal transmission,  exceptional points, and topological phase transitions \citep{Exceptional.topology.RevModPhys2021,exceptional-point.geometries2022}.
These phenomena have attracted significant attention in recent years due to their potential applications in various fields of physics, such as optics, condensed matter physics, and quantum information science.

With the non-Hermitian and topological nature, we foresee that the rotating convection will become a more abundant system and act as a platform for probing topological states in turbulent flows. This paper is structured as follows. Sec.\ref{sec:Non-Hermitian Hamiltonian matrix} introduces the Non-Hermitian Hamiltonian matrix of the linearized governing equations. Sec.\ref{Chern number in non-Hermitian} explains the calculation of Chern number in non-Hermitian systems. Sec.\ref{unstable} focuses on calculating the topological invariants in the unstably stratified case. Sec.\ref{stable_critical} explores the topological properties of the system in stable and critical cases. Sec.\ref{Hydrostatic} discusses the calculation of the Chern number under the hydrostatic approximation. 
Sec.\ref{Winding} visualizes the topological nature through winding numbers.
Finally, discussions and summaries are presented in Sec. \ref{discussion&summary}. 
The appendix Sec. \ref{appA} discusses the solutions satisfying realistic boundary conditions.

\section{The non-Hermitian eigenvalue problem of linearized rotating convection}

\subsection{Non-Hermitian Hamiltonian matrix}
\label{sec:Non-Hermitian Hamiltonian matrix}

%In convection, the fluid motion is driven by buoyancy and rotation and is characterized by the Rayleigh number ($Ra$), the Ekman number ($E$), and the Prandtl number ($Pr$). 
%Friction and heat diffusion are described in terms of constant viscosity ($\nu$) and thermal diffusivity ($\kappa$).
The governing equations for the rotating Rayleigh-B\'enard convection include the Navier-Stokes, continuity, and temperature equations. The dimensionless form can be written as \cite[cf.][]{Favier_Knobloch.2020}:
\begin{subequations}\label{RRBC}
	\begin{align}
		\frac{\partial\boldsymbol{u}}{\partial t}+\left(\boldsymbol{u}\cdot\boldsymbol{\nabla}\right)\boldsymbol{u}&=-\boldsymbol{\nabla}p-\lambda\boldsymbol{e}_{z}\times\boldsymbol{u}+\alpha\theta\boldsymbol{e}_{z}+E\nabla^{2}\boldsymbol{u}, \\
		\boldsymbol{\nabla}\cdot\boldsymbol{u}&=0,\\
		\frac{\partial\theta}{\partial t}+\left(\boldsymbol{u}\cdot\boldsymbol{\nabla}\right)\theta&=w+\frac{E}{Pr}\nabla^{2}\theta,
	\end{align}
\end{subequations}
where $\boldsymbol{u}=(u,v,w)$ is the velocity vector, $\lambda=\pm1$ indicates the direction of rotation,
$p$ is the pressure, $\theta$ is the temperature fluctuation relative to the
linear conduction profile, $\alpha=RaE^{2}/Pr$ is the square of the convective Rossby number. 
The Rayleigh number is $Ra=\beta g\Delta Th^{3}/\nu\kappa$,
where $\beta$ is the thermal expansion coefficient, $g$ is the acceleration due to gravity, $\Delta T$ is the temperature difference between the top and bottom plates, $h$ is the height of the fluid layer, and $\nu$ and $\kappa$ are the viscosity and thermal diffusivity, respectively. 
It is a measure of the strength of the buoyancy-driven flow relative to the viscous forces. 
The Ekman number $E =\nu/(2\Omega h^2)$, describes the balance of viscous forces to Coriolis forces, where $\Omega$ is the angular velocity of rotation. 
The Prandtl number is $Pr=\nu/\kappa$ and is a measure of the relative importance of viscous and thermal diffusion.

Considering a normal-mode ansatz $(u,v,w,p,\theta)=Re\{(\hu,\hv,\hw,\hp,\htheta) e^{i(\boldsymbol{k\cdot r}-\omega t)}\}$ with the wavenumber $\boldsymbol{k}=(k_x,k_y,k_z)$, the linearized equations become
\begin{equation}
-i\omega\boldsymbol{\hu}=-i\hp\boldsymbol{k}-\lambda\boldsymbol{e}_{z}\times\boldsymbol{\hu}+\alpha\htheta\boldsymbol{e}_z-Ek^{2}\boldsymbol{\hu},\label{eq:eqk1}
\end{equation}
\begin{equation}
\boldsymbol{k}\cdot\boldsymbol{\hu}=0,\label{eq:eqk2}
\end{equation}
\begin{equation}
-i\omega\htheta=\hw-\frac{E}{Pr}k^{2}\htheta,\label{eq:eqk3}
\end{equation}
where $k=\sqrt{k_{x}^{2}+k_{y}^{2}+k_{z}^{2}}$.
Taking the divergence of both sides of Eq.(\ref{eq:eqk1})
and combining with Eq.(\ref{eq:eqk2}), we get
\begin{equation}
0=\hp k^{2}+i\lambda(k_{x}\hv-k_{y}\hu)+i\alpha k_{z}\htheta.
\end{equation}
Then
\begin{equation}
\hp=i\frac{\lambda(-k_{x}\hv+k_{y}\hu)-\alpha k_{z}\htheta}{k^{2}}.
\end{equation}

With the wave vector $\psi\equiv(\hu,\hv,\hw,\htheta)$, we get the eigen equation from Eq.(\ref{eq:eqk1}) and Eq.(\ref{eq:eqk3}) as
\begin{equation}\label{system}
	H\psi=\omega\psi,
\end{equation}
with
\begin{equation}
	H=i\left[\begin{array}{cccc}
		E_{0}+\lambda\frac{k_{x}k_{y}}{k^{2}} & \lambda\left(1-\frac{k_{x}^{2}}{k^{2}}\right) & 0 & -\alpha\frac{k_{x}k_{z}}{k^{2}}\\
		\lambda\left(-1+\frac{k_{y}^{2}}{k^{2}}\right) & E_{0}-\lambda\frac{k_{x}k_{y}}{k^{2}} & 0 & -\alpha\frac{k_{y}k_{z}}{k^{2}}\\
		\lambda\frac{k_{y}k_{z}}{k^{2}} & -\lambda\frac{k_{x}k_{z}}{k^{2}} & E_{0} & \alpha-\alpha\frac{k_{z}^{2}}{k^{2}}\\
		0 & 0 & 1 & E_{1}
	\end{array}\right],
\end{equation}
where $E_{0}=-Ek^{2}$, $E_{1}=E_{0}/Pr$. This is a non-Hermitian Hamiltonian because $H\neq H^{\dagger}$, where $\dagger$ denotes the conjugate transpose. The eigenvalue $\omega$ can take a complex number, unlike in the Hermitian case where $\omega$ is real.
The eigenvalue problem of $H^{\dagger}$ is
\begin{equation}
	H^{\dagger}\psi^{\prime}=\omega^{\ast}\psi^{\prime},
\end{equation}
where $\omega^{\ast}$ is the complex conjugate of $\omega$, and $\psi^{\prime}$ is generally different from $\psi$.

In (\ref{RRBC}), we do not include the boundary conditions, and the main text of this paper focuses on demonstrating the idea that to calculate a Chern number for rotating convection a complex vertical wavenumber is required.
A discussion on realistic boundary conditions can be found in \S \ref{appA}.

\subsection{Chern number in a non-Hermitian system}
\label{Chern number in non-Hermitian}

Chern number is a topological invariant that was originally defined for systems with a periodic structure, such as crystals, by integrating the Berry curvature over the Brillouin zone \citep{zak1989,xiao2010}. 
It is a quantity with integer values and is numerically equal to the Berry phase divided by $2\pi$.
The Berry curvature characterizes the local geometry of wave polarization and is recognized to manifest in the equations of motion of wave packets with multiple components \citep{berry1984,Perez_et_al.2021}.

As a topological invariant, the Chern number is insensitive
to small perturbations of the Hamiltonian that do not change the topology of the system. 
A non-zero Chern number implies the existence of a nontrivial bulk topology and the presence of robust edge states \citep{TIs2010}.
The bulk-boundary correspondence principle links the topological properties of the bulk and edge states \citep{PRB2011,raytracing2023}.

In the wavenumber space, the Berry connection \citep{berry1984} is defined as
\begin{equation}
	A_{n}(\boldsymbol{k})=i\langle \psi_{n}(\boldsymbol{k})|\nabla_{\boldsymbol{k}}|\psi_{n}(\boldsymbol{k})\rangle,
\end{equation}
where $\psi_{n}$ is the periodic part of the Bloch wave function $\phi_{n}=\psi_{n}e^{i\boldsymbol{k\cdot r}}$, and $\nabla_{\boldsymbol{k}}$ is the gradient with respect to the wave vector $\boldsymbol{k}$. 
As the mathematical symbols of quantum mechanics, the right vector $|\psi_{n}\rangle$ represents the general eigenvector, the left vector $\langle\psi_{n}|$ represents the conjugate transpose of $|\psi_{n}\rangle$, and $\langle\cdots|\cdots\rangle$ represents the inner product.
The Berry connection is a vector-valued function whose integral along a closed path gives the Berry phase. 
Then, the Berry curvature is calculated by the curl of the Berry connection 
\begin{equation}
	\boldsymbol{\Omega}_{n}(\boldsymbol{k})=\nabla_{\boldsymbol{k}}\times\boldsymbol{A}_{n}(\boldsymbol{k}).
\end{equation}
The Chern number is then defined as the integral of the Berry curvature over the Brillouin zone:
\begin{equation}
	C_{n}=\frac{1}{2\pi}\int_{\text{BZ}}\boldsymbol{\Omega}_{n}(\boldsymbol{k})\cdot d\boldsymbol{S},
\end{equation}
where $d\boldsymbol{S}$ is the surface element of the Brillouin zone. The Berry curvature $\boldsymbol{\Omega}_{n}$ is a pseudovector in three dimensions. Due to the direction of rotation in this system, we only consider its component in the $z$ direction, $\varOmega_{n}^{z}$, and $d\boldsymbol{S}$ takes the horizontal plane.

In non-Hermitian systems, the classic Brillouin zone is no longer sufficient to describe the band topology due to the significant difference in frequency spectra between open (non-Bloch) and periodic (Bloch and classic) boundaries. Therefore, a generalized Brillouin zone (GBZ) defined on the complex wavenumber space is needed.
As a result, a non-Bloch Chern number is defined as the integral of the Berry curvature over the generalized Brillouin zone:
\begin{equation}
	C_{n}=\frac{1}{2\pi}\int_{\text{GBZ}}\boldsymbol{\Omega}_{n}(\tilde{\boldsymbol{k}})\cdot d\tilde{\boldsymbol{S}},
\end{equation}
where $\tilde{\boldsymbol{k}}$ is complex and $d\tilde{\boldsymbol{S}}$ is the surface element of the generalized Brillouin zone. 
The Berry curvature can take a complex number, as long as the Chern number obtained by integration remains an integer \citep{Non-Hermitian.Chern.Bands}.

Due to the non-Hermitian nature of the system (\ref{system}), obtaining the Berry curvature requires the consideration of a biorthogonal basis set, $\{|\psi_{n}\rangle\}$ and $\{|\psi^{\prime}_{n}\rangle\}$, which satisfy
\begin{subequations}
\begin{align}
H|\psi_{n}\rangle&=\omega_{n}|\psi_{n}\rangle,\\
H^{\dagger}|\psi^{\prime}_{m}\rangle&=\omega_{m}^{\ast}|\psi^{\prime}_{m}\rangle,\\
\langle\psi^{\prime}_{m}|\psi_{n}\rangle&=\delta_{m,n}, \label{orth}
\end{align}
\end{subequations}
where $\omega_m$ and $\omega_n$ are eigenvalues with band indexes $m$ and $n$.
Then we can define the non-Hermitian Berry curvature of band $\omega_{n}$ as  \citep{Non-Hermitian.Chern.Bands}:
\begin{equation}\label{berry_curv}
\varOmega_{n}^{z}=i\left[\left\langle \frac{\partial\psi_{n}^{\prime}}{\partial k_{x}}\left|\frac{\partial\psi_{n}}{\partial k_{y}}\right.\right\rangle -\left\langle \frac{\partial\psi_{n}^{\prime}}{\partial k_{y}}\left|\frac{\partial\psi_{n}}{\partial k_{x}}\right.\right\rangle \right].
\end{equation}

\section{Topology of rotating convection}
\label{unstable}

\subsection {Complex wave number $k_{z}$ }

When the eigenvalues of the linearized equations Eqs.(\ref{eq:eqk1})-(\ref{eq:eqk3}) are not real, which often happens in non-Hermitian systems, the convection is linearly unstable with oscillatory modes that can grow or decay over time, and exploring the topological properties gets complicated.
For example, in the inviscid case where $E=0$, the eigenvalues and eigenvectors are degenerated at $k_{x}^{2}+k_{y}^{2}=k_{z}^{2}/\alpha$ (cf. Eq.(\ref{eigenvalues})), and a larger $k_{x}$ or  $k_{y}$ would make the eigenvalues purely imaginary.
\citet{prl.120.146402} proposed that the non-Hermitian system loses its well-defined Chern numbers unless $k_{z}$ becomes complex: with a non-zero imaginary part in $k_{z}$, the frequency bands will always remain gapped in the complex space, and the integration of the Berry curvature will give an integer Chern number.

\begin{figure}
	\centering
	\includegraphics[width=0.6\linewidth]{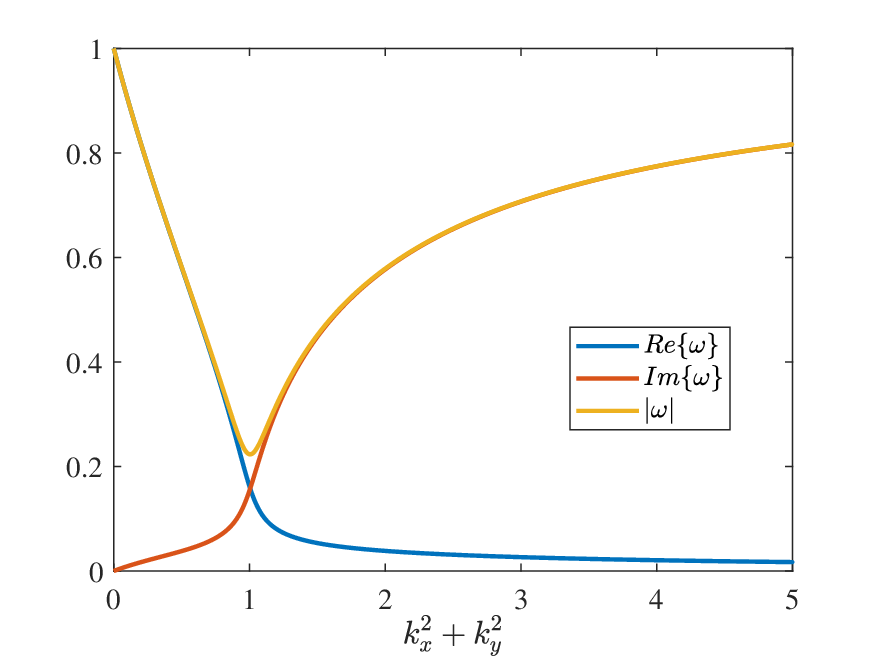}
	\caption{Illustration of the eigenvalue with complex $k_z$.
		Here, $k_z^2=1+0.1i$, $\alpha=1$ and $E=0$.
	}
	\label{fig:gap}
\end{figure}

For the rotating convection system, Figure \ref{fig:gap} illustrates the eigen frequency gap opened by introducing a complex vertical wavenumber $k_z$ with $k_z^2=1+0.1i$ and $\alpha=1$.
We can see that both the real and imaginary parts of $\omega$ approach $0$, but its absolute value is bounded from below.
When $k_z^2=1+\epsilon i$ with $\epsilon\ll1$, one can asymptotically find that
\begin{equation}
  |\omega|_{gap} = \sqrt{\frac{\epsilon\alpha}{1+\alpha}} + h.o.t.,
\end{equation}
where $h.o.t.$ denotes high-order terms.

Complex $k_z$ is also physically reasonable.
The top and bottom non-slip boundaries introduce an exponential decay form to the asymptotic solutions along the $z$-direction (\citet{Zhang_Liao.2009}), which agrees with our numerical results in Appendix \ref{appA}.
Also, under open boundary conditions, i.e., with external energy injection, the bulk eigen states of non-Hermitian systems exhibit a localized behavior towards the border, the so-called non-Hermitian skin effect, which differs from the extended Bloch waves in Hermitian systems.
In our system, energy exchanges with the outsiders in the $z$-direction due to the temperature difference, so we take a complex $k_{z}$.

\subsection{Inviscid topological invariant ($E=0$)}
\label{inviscid}

With the generalized Brillouin zone where $k_{z}$ is complex, we can calculate the Chern number for open boundaries through a normal process, just as we do under periodic boundary conditions.
The solutions satisfying realistic boundary conditions can be found in Appendix \ref{appA}.
To obtain a single-valued function when dealing with the square root, 
we redefine the square root operation on the complex domain, making its value domain lie in the right half-plane that does not include the negative imaginary axis, i.e., $\sqrt{-1}=i$.

We start to discuss the topological properties of the system with the simplest case $E=0$. Since $k_{z}$ is complex, the singularity in the eigenvalues and eigenvectors will not exist. The Hamiltonian matrix becomes
\begin{equation}
H_{0}=i\left[\begin{array}{cccc}
\lambda\frac{k_{x}k_{y}}{k^{2}} & \lambda(1-\frac{k_{x}^{2}}{k^{2}}) & 0 & -\alpha\frac{k_{x}k_{z}}{k^{2}}\\
\lambda(-1+\frac{k_{y}^{2}}{k^{2}}) & -\lambda\frac{k_{x}k_{y}}{k^{2}} & 0 & -\alpha\frac{k_{y}k_{z}}{k^{2}}\\
\lambda\frac{k_{y}k_{z}}{k^{2}} & -\lambda\frac{k_{x}k_{z}}{k^{2}} & 0 & \alpha-\alpha\frac{k_{z}^{2}}{k^{2}}\\
0 & 0 & 1 & 0
\end{array}\right].
\end{equation}
 For the eigen equation
\begin{equation}
H_{0}\psi=\omega\psi,
\end{equation}
the non-zero eigenvalues and corresponding right eigenvectors are
\begin{equation}
\omega_{\pm}=\pm\omega_{0},\ \ \omega_{0}=\frac{\sqrt{k_{z}^{2}-\alpha(k_{x}^{2}+k_{y}^{2})}}{k},
\label{eigenvalues}
\end{equation}
\begin{equation}
\left.\left|\psi_{\pm}\right.\right\rangle =\frac{1}{\sqrt{2\left(\frac{k_{z}^{2}}{k_{x}^{2}+k_{y}^{2}}-\alpha\right)}}\left[\begin{array}{c}
\frac{k_{z}\left(-i\lambda k_{y}\mp k_{x}\omega_{0}\right)}{k_{x}^{2}+k_{y}^{2}}\\
\frac{k_{z}\left(i\lambda k_{x}\mp k_{y}\omega_{0}\right)}{k_{x}^{2}+k_{y}^{2}}\\
\pm\omega_{0}\\
i
\end{array}\right],\label{eq:rightvec}
\end{equation}
where $k=\sqrt{k_x^2+k_y^2+k_z^2}$ and  $k_x^2+k_y^2\neq 0$. When $\rho=\sqrt{k_{x}^{2}+k_{y}^{2}}\xrightarrow{}0$, the eigenvectors become
\begin{equation}
\left.\left|\psi_{\pm}\right.\right\rangle =\frac{1}{\sqrt{2}}\left[\begin{array}{c}
\pm \lambda i\\
1\\
0\\
0
\end{array}\right].
\end{equation}
When $\rho\xrightarrow{}+\infty$, the eigenvectors are still single-valued that
\begin{equation}
\left.\left|\psi_{\pm}\right.\right\rangle =\frac{1}{\sqrt{2}}\left[\begin{array}{c}
0\\
0\\
\pm 1\\
\frac{i}{\sqrt{-\alpha}}
\end{array}\right].
\end{equation}
Thus, different from the shallow water model, the eigenvectors here are regular on a compact manifold, which guarantees a well-defined Chern number without introducing unphysical items \citep{tauber2019}.

For the eigen equation
\begin{equation}
H_{0}^{\dagger}\psi^{\prime}=\omega^{\ast}\psi^{\prime},
\end{equation}
the non-zero eigenvalues and corresponding left eigenvectors are
\begin{equation}
\omega_{\pm}^{\ast}=\pm\omega_{0}^{\ast},
\end{equation}
\begin{equation}
\left\langle \left.\psi_{\pm}^{\prime}\right|\right.=\frac{1}{\sqrt{2\left(\frac{k_{z}^{2}}{k_{x}^{2}+k_{y}^{2}}-\alpha\right)}}\left[\begin{array}{c}
\frac{k_{z}\left(i\lambda k_{y}\mp\frac{k_{x}}{\omega_{0}}\right)}{k_{x}^{2}+k_{y}^{2}}\\
\frac{k_{z}\left(-i\lambda k_{x}\mp\frac{k_{y}}{\omega_{0}}\right)}{k_{x}^{2}+k_{y}^{2}}\\
\mp\frac{\alpha}{\omega_{0}}\\
i\alpha
\end{array}\right]^{\mathrm{T}}, \label{eq:leftvec}
\end{equation}
when $k_x^2+k_y^2\neq 0$.
When $\rho\xrightarrow{}0$, the eigenvectors become
\begin{equation}
\left\langle \left.\psi_{\pm}^{\prime}\right|\right. =\frac{1}{\sqrt{2}}\left[\begin{array}{c}
\mp \lambda i\\
1\\
0\\
0
\end{array}\right]^{\mathrm{T}}.
\end{equation}
When $\rho\xrightarrow{}+\infty$, the eigenvectors are single-valued that
\begin{equation}
\left\langle \left.\psi_{\pm}^{\prime}\right|\right. =\frac{1}{\sqrt{2}}\left[\begin{array}{c}
0\\
0\\
\pm 1\\
-i\sqrt{-\alpha}
\end{array}\right]^{\mathrm{T}}.
\end{equation}

Substituting the expressions of eigenvectors in Eq.(\ref{berry_curv}), the Berry curvature of the positive band (associated with $\omega_{+}$ and $\omega_{+}^{\ast}$) becomes
\begin{equation}
	\varOmega^{z}=-\lambda\frac{k_{z}^{2}\left[\rho^{2}k_{z}^{2}+2\alpha\left(\rho^{4}+4\rho^{2}k_{z}^{2}+2k_{z}^{4}\right)-\alpha^{2}\rho^{2}\left(2\rho^{2}+k_{z}^{2}\right)\right]}{2k^{3}\left(k_{z}^{2}-\alpha\rho^{2}\right)^{5/2}},\label{eq:BerryCur}
\end{equation}
where $\rho=\sqrt{k_{x}^{2}+k_{y}^{2}}$.
Then the Chern number is
\begin{subequations}
	\begin{align}
		C&=\frac{1}{2\pi}\int_{-\infty}^{+\infty}\int_{-\infty}^{+\infty}\varOmega^{z}dk_{x}dk_{y}\\
		&=\int_{0}^{+\infty}\rho\varOmega^{z}d\rho\\
		&=\lambda\left.\frac{-2k_{z}^{4}+k_{z}^{2}(\alpha-1)\rho^{2}}{2\sqrt{\rho^{2}+k_{z}^{2}}\left(k_{z}^{2}-\alpha\rho^{2}\right)^{3/2}}\right|_0^{+\infty}\\
		&=\lambda,
	\end{align} 
\end{subequations}
which implies the existence of topologically protected edge states only related to the direction of fluid rotation.
This is a three-dimensional version of the topologically protected equatorial waves \citep{Delplace2017}. 
Under the hydrostatic approximation, we obtain an exact two-dimensional counterpart, shown in Sec.\ref{Hydrostatic}.

\subsection {Topological invariant when $E\neq0$}

This section considers $E\neq0$, which does not change the Chern number we obtained in the previous subsection.
Firstly, When $Pr=1$, the Hamiltonian matrix $H$ becomes
\begin{equation}\label{H,Pr=1}
	H=H_{0}+iE_{0}\boldsymbol{\hat{1}},
\end{equation}
where $H_{0}$ is the simple  Hamiltonian where  $E=0$  and $\boldsymbol{\hat{1}}$ is the unit matrix of the same order as $H_{0}$. The eigenvalues are 
\begin{equation}\label{omega,Pr=1}
	\omega_{\pm}=\pm\omega_{0}+iE_{0}.
\end{equation}
It is just a shift of $\pm\omega_{0}$, the eigenvalues of $H_{0}$, and then the eigenvectors are independent of $E$. Therefore, both the Berry curvature and Chern number are unchanged compared with the case where $E=0$.

\begin{figure}
	\centering
	\includegraphics[width=0.8\linewidth]{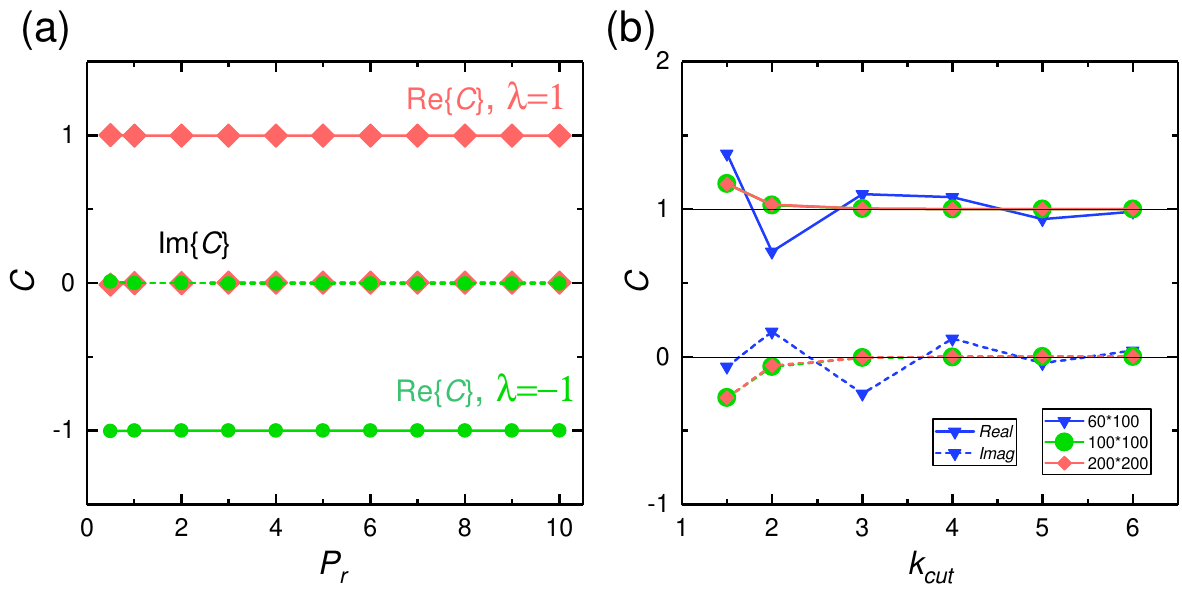}
	\caption{Numerical results of the non-Hermitian Chern number.
		(a) Chern number with different Prandtl number $Pr$ and rotation directions $\lambda=\pm1$.
		The upper limit of integral for $\rho$ is $k_{cut}=6$, the $k$-point accuracy per unit length is 200 (radial) * 200 (axial) using the Gauss-Legendre integration scheme.
		(b) Demonstrating the numerical convergence degree with different $k$-point accuracy as $k_{cut}$ increases.
		Other parameters are $\alpha=1$, $E=0.01$, $k_{z}=1+0.1 i$.
	}
	\label{fig:Pr}
\end{figure}

When $Pr\neq1$, there are no simple expressions for the eigenvalues and eigenvectors, and we need to calculate the Chern number with the help of numerical calculations, seen in Figure \ref{fig:Pr}. 
To avoid the derivation of the eigenvectors, we rewrite Eq.(\ref{berry_curv}) as
\begin{equation}
	\Omega_{n}^{z}=i\underset{m\neq n}{\sum}\frac{<\psi_{n}^{\prime}|\frac{\partial H}{\partial k_{x}}|\psi_{m}><\psi_{m}^{\prime}|\frac{\partial H}{\partial k_{y}}|\psi_{n}>-\{x\leftrightarrow y\}}{(\omega_{n}-\omega_{m})^{2}}.
\end{equation}
It is similar to the Hermitian case \citep{xiao2010} except that the left vector takes the biorthogonal partner. This form is very useful for numerical calculations because it can be evaluated under an unsmooth phase choice of the eigenstates, which often occurs in the standard diagonalization algorithms.

\section{The stably stratified and critical cases}
\label{stable_critical}

Even though this paper focuses on the unstably stratified convection situation, this section shows that following the above-mentioned procedure for calculating the Chern number in non-Hermitian systems we can recover the classic results in the stably stratified situations \cite[e.g.][]{Delplace2017}.
For the simple case that  $E=0$, when $\alpha<0$, the eigenvalues of the linearized equations Eqs. (\ref{eq:eqk1})-(\ref{eq:eqk3}) are real, and the system is stable without oscillatory modes that can grow or decay over time. 
At this time the discussion in the previous section where  $\alpha>0$ still holds, and the system still has a non-zero Chern number $C=\lambda$. Thus, the existence of the tenacious wall modes in the stratified fluid relies only on rotation, regardless of the driving mechanism.

When $\alpha=0$, the system is in a critical neutral state, and the eigenvalues of the Hamiltonian are 
\begin{subequations}\label{critical}
\begin{align}
    \omega_{1}\equiv\omega_{+}&=\frac{k_{z}}{k}-iEk^{2},\\
    \omega_{2}\equiv\omega_{-}&=-\frac{k_{z}}{k}-iEk^{2},\\
    \omega_{3}&=-i\frac{E}{Pr}k^{2},\\
    \quad\omega_{4}&=-iEk^{2}.
\end{align}    
\end{subequations}
Considering the symmetry between $\omega_{+}$ and $\omega_{-}$, we have assumed that $Re\{k_{z}\}>0$.
The Berry curvature of band $\omega_{3}$ (or $\omega_{4}$) is 0.
The Berry curvature of band $\omega_{\pm}$ reads
\begin{equation}
\varOmega^{z}_{\pm}=\mp\lambda\frac{k_{x}^{2}+k_{y}^{2}}{2k_{z}k^{3}},
\end{equation}
which is independent of $E$ and $Pr$. The integral does not converge 
when we evaluate the Chern number
\begin{equation}
C_{\pm}=\frac{1}{2\pi}\int_{-\infty}^{+\infty}\int_{-\infty}^{+\infty}\varOmega^{z}_{\pm}dk_{x}dk_{y}
=\int_{0}^{+\infty}\frac{\mp\lambda\rho^{3}}{2k_{z}(\rho^2+k_{z}^2)^{3/2}}d\rho.
\end{equation}
Mathematically, this is due to the degeneration of $\omega_{+}$ and $\omega_{-}$ at $\rho\xrightarrow{}+\infty$, which leads to the close of the band gap of the system.
The critical case is important to show that stratification is crucial, whether stable or unstable.

\section{Hydrostatic approximation}
\label{Hydrostatic}

Under the hydrostatic approximation, the $z$-direction momentum equation of (\ref{eq:eqk1}) reduces to
\begin{equation}
	\htheta=\frac{i\hp k_{z}}{\alpha}.
\end{equation}

From Eq.(\ref{eq:eqk2}) we get
\begin{equation}
	\hw=-\frac{k_{x}\hu+k_{y}\hv}{k_{z}}.
\end{equation}

With the wave vector $\psi\equiv(\hu,\hv,\hp)$, we get the eigen equations
from Eq.(\ref{eq:eqk1}) and Eq.(\ref{eq:eqk3}) as
\begin{equation}
	H=\left[\begin{array}{ccc}
		iE_{0} & i\lambda & k_{x}\\
		-i\lambda & iE_{0} & k_{y}\\
		-\alpha\frac{k_{x}}{k_{z}^{2}} & -\alpha\frac{k_{y}}{k_{z}^{2}} & iE_{1}
	\end{array}\right],
\end{equation}
where $E_{0}=-Ek^{2}$, $E_{1}=E_{0}/Pr$. This is a non-Hermitian Hamiltonian because $H\neq H^{\dagger}$, but for the stable case where $\alpha<0$, it can be changed into a Hermitian one by replacing the variables. 
If we further assume that $Pr=1$, the Hamiltonian is an analogue of the one in topological shallow water waves in \cite{Delplace2017}, except for a frequency shift.
However, for the unstable case with  $\alpha>0$, the non-Hermitian nature of the system is intrinsic and cannot be turned into a Hermitian one by variable substitutions.

For the eigen equation
\begin{equation}
	H\psi=\omega\psi,
\end{equation}
the eigenvalues and eigenvectors contributing to non-zero Chen numbers are ($Pr=1$)
\begin{equation}
	\omega_{\pm}=\pm\omega_{0}-iEk^{2},\ \ \omega_{0}=\sqrt{1-\frac{\alpha}{k_{z}^{2}}(k_{x}^{2}+k_{y}^{2})},
\end{equation}
\begin{equation}
	\left.\left|\psi_{\pm}\right.\right\rangle =\frac{1}{\sqrt{2\left(\frac{k_{z}^{2}}{k_{x}^{2}+k_{y}^{2}}-\alpha\right)}}\left[\begin{array}{c}
		\frac{k_{z}^{2}\left(i\lambda k_{y}\pm k_{x}\omega_{0}\right)}{k_{x}^{2}+k_{y}^{2}}\\
		\frac{k_{z}^{2}\left(-i\lambda k_{x}\pm k_{y}\omega_{0}\right)}{k_{x}^{2}+k_{y}^{2}}\\
		-\alpha
	\end{array}\right].\label{eq:eigenvector_in_static}
\end{equation}
When $\rho\xrightarrow{}0$, the eigenvectors become
\begin{equation}
	\left.\left|\psi_{\pm}\right.\right\rangle =\frac{1}{\sqrt{2}}\left[\begin{array}{c}
		\pm \lambda i\\
		1\\
		0
	\end{array}\right].
\end{equation}
When $\rho\xrightarrow{}+\infty$, the eigenvectors become
\begin{equation}
	\left.\left|\psi_{\pm}\right.\right\rangle =\frac{1}{\sqrt{2}}\left[\begin{array}{c}
		\pm \sqrt{k_z^2}\frac{k_x}{\rho}\\
		\pm \sqrt{k_z^2}\frac{k_y}{\rho}\\
		\sqrt{-\alpha}
	\end{array}\right],
\end{equation}
which are not single-valued in different directions. In order to get a compact manifold where the Chern number is well defined, we may need to introduce an odd viscosity term like in the shallow water model \citep{tauber2019}.

For the eigen equation
\begin{equation}
H^{\dagger}\psi^{\prime}=\omega^{\prime}\psi^{\prime},
\end{equation}
we have
\begin{equation}
\omega_{\pm}^{\prime}=\omega_{\pm}^{\ast},
\end{equation}
\begin{equation}
\left\langle \left.\psi_{\pm}^{\prime}\right|\right.=\frac{1}{\sqrt{2\left(\frac{k_{z}^{2}}{k_{x}^{2}+k_{y}^{2}}-\alpha\right)}}\left[\begin{array}{c}
\frac{-i\lambda k_{y}\pm k_{x}\omega_{0}}{k_{x}^{2}+k_{y}^{2}}\\
\frac{i\lambda k_{x}\pm k_{y}\omega_{0}}{k_{x}^{2}+k_{y}^{2}}\\
1
\end{array}\right]^{\mathrm{T}}.\label{eq:left_in_static}
\end{equation}
When $\rho\xrightarrow{}0$, the eigenvectors become
\begin{equation}
\left\langle \left.\psi_{\pm}^{\prime}\right|\right. =\frac{1}{\sqrt{2}}\left[\begin{array}{c}
\mp \lambda i\\
1\\
0
\end{array}\right]^{\mathrm{T}}.
\end{equation}
When $\rho\xrightarrow{}+\infty$, the eigenvectors become
\begin{equation}
	\left\langle \left.\psi_{\pm}^{\prime}\right|\right. =\frac{1}{\sqrt{2}}\left[\begin{array}{c}
		\pm \frac{k_x}{\rho\sqrt{k_z^2}}\\
		\pm \frac{k_y}{\rho\sqrt{k_z^2}}\\
		\frac{1}{\sqrt{-\alpha}}
	\end{array}\right].
\end{equation}

The Berry curvature of the positive band (associated with $\omega_{+}$ and $\omega_{+}^{\prime}$)
is
\begin{equation}
\varOmega^{z}=-\frac{\lambda\alpha\sqrt{k_{z}^{2}}}{\left(k_{z}^{2}-\alpha\rho^{2}\right)^{3/2}},
\end{equation}
where $\rho=\sqrt{k_{x}^{2}+k_{y}^{2}}$.
Then the Chern number is
\begin{subequations}
   \begin{align}
       C&=\frac{1}{2\pi}\int_{-\infty}^{+\infty}\int_{-\infty}^{+\infty}\varOmega^{z}dk_{x}dk_{y}\\
&=\int_{0}^{+\infty}\rho\varOmega^{z}d\rho\\
&=\left.-\frac{\lambda\sqrt{k_{z}^{2}}}{\sqrt{k_{z}^{2}-\alpha\rho^{2}}}\right|_0^{+\infty}\\
&=\lambda. 
\end{align} 
\end{subequations}
In the stably stratified case, $\alpha<0$, the above calculation holds when $Re\{k_{z}\}\neq0$;
while in the unstably stratified case, $\alpha>0$, we need $Im\{k_{z}\}\neq0$.
If the system is neutral, $\alpha=0$, we arrive at two gapped flat bands $\omega_{\pm}=\pm1-iEk^{2}$, which differ from the degenerate scenario in Eq.(\ref{critical}), and obtain $C=0$ since $\varOmega^{z}=0$.

\section{Winding number}
\label{Winding}

\begin{figure}
	\centering
	\includegraphics[width=0.45\linewidth]{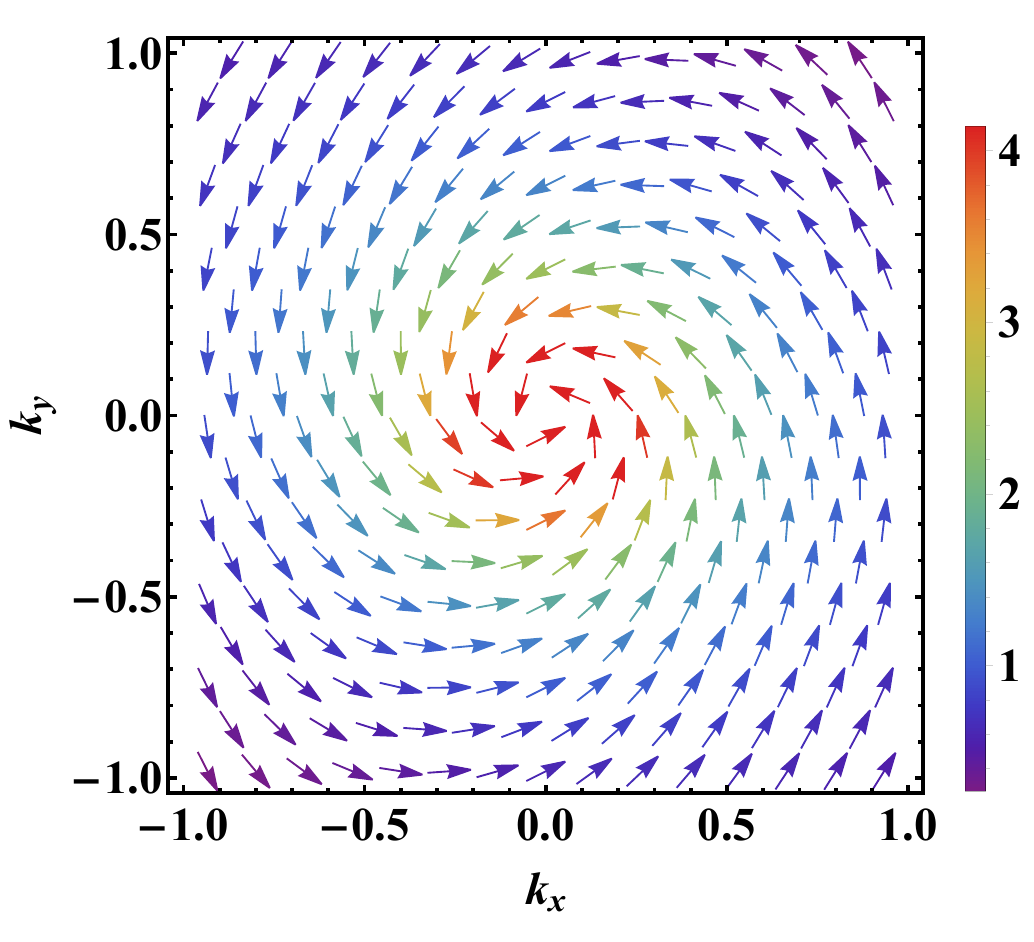}
        \includegraphics[width=0.45\linewidth]{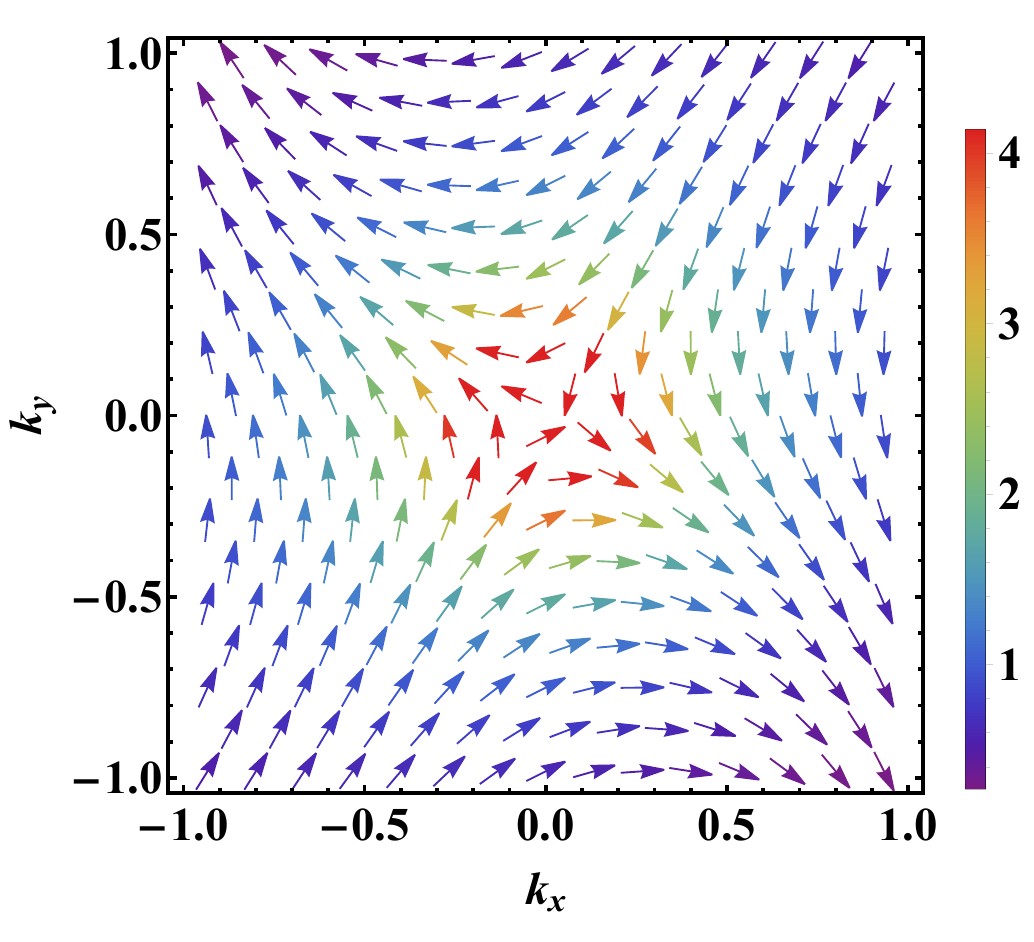}
	\caption{Arrows representing argument of $\Xi(\boldsymbol{k})=\hv(\boldsymbol{k})*\hw^{\ast}(\boldsymbol{k})$ with $\lambda=1$ (left) and $\lambda=-1$ (right). The x and y components of the arrow represent the real and imaginary parts of $\Xi(\boldsymbol{k})$. The length of the arrows is rescaled to be equal. Colors represent normalized magnitude $|\Xi(\boldsymbol{k})|$ in arbitrary units.
		Other parameters are $Pr=1$, $\alpha=0.5$,  $k_{z}=1+0.5 i$.
	}
	\label{fig:Winding}
\end{figure}

To put the topological nature of the system into perspective, we define a complex function in the wavenumber space $(k_x,k_y)$ \cite[cf.][]{winding_number2023}:
\begin{equation}
\Xi(\boldsymbol{k})=\hv(\boldsymbol{k})\hw^{\ast}(\boldsymbol{k}),
\end{equation}
which removes the gauge redundancy of the eigenfunctions, leaving only the internal phase difference between the two components.
 Figure \ref{fig:Winding} depicts the argument of $\Xi(\boldsymbol{k})$, $\tan^{-1}(Re\{\Xi\}/Im\{\Xi\})$, whose x and y components represent the real and imaginary parts of $\Xi$ with a rescaled equal length.
When $\lambda=1$, there is a vortex formed by the vector arrows going in the anticlockwise direction, and the arrows along a closed circle smoothly wind by a phase of 2$\pi$, suggesting a winding number of $1$.
Conversely, when $\lambda=-1$, there is a strain flow with the winding number of $-1$.

\section{Summary and Discussion}
\label{discussion&summary}

In summary, we show that the linearized rotating Rayleigh-B\'enard convection can support non-Hermitian topological states characterized by a non-zero integer Chern number. 
Thus, we verify the hypothesis by \citet{Favier_Knobloch.2020} that the robust wall modes in rapidly rotating convection are topologically protected.
Due to the unstable stratification, the linear eigenvalue problem is intrinsically non-Hermitian, so the Berry curvature is defined on biorthogonal eigenstates, and the corresponding eigenvalues may be complex, which is very different from the Hermitian system. Beyond the shallow water model, the eigenvectors here are regular on a compact manifold, which guarantees a well-defined Chern number without introducing unphysical items \citep{tauber2019}.
The emergence of these topological edge states is fundamentally due to rotation breaking the system's time-reversal symmetry, but stratification also plays a crucial role: without the stratification,  i.e., for the critical case, the bulk Chern number is either not well-defined or equals zero.
Under the hydrostatic approximation, the problem transforms into a two-dimensional counterpart of the one that explains the topological origin of equatorial waves \citep{Delplace2017}.
Finally, an eigenvector-dependent winding number is introduced to visualize the topological nature of the fluid.

On the precession direction of the edge states, they are usually (not all) retrograde in rotating convection \citep{PhysRevLett.1991Asymmetric,goldstein_knobloch_mercader_net_1993,PhysRevE.47.R2245,herrmann_busse_1993}, but prograde both at the Earth's equator and in the two-dimensional fluids with odd viscosity \citep{Delplace2017,odd2019prl}.
This is a result of the combined contribution of the boundary conditions and the system parameters, especially the effect of the Prandtl number $Pr$ \citep{goldstein_1994,horn_schmid_2017}.
It is quite different from the conventional Hermitian system with periodic boundary conditions, where the positive or negative sign of the Chen number determines the direction of the edge states \citep{TIs2010,PRB2011}.

For simplicity, we assume both $k_{x}$ and $k_{y}$ to be real numbers. 
Thus, according to the regular formalities \citep{odd2019prl}, the non-zero Chern number we obtain describes the topological properties of the system without boundary in the $x-y$ plane.
When there is an impenetrable and slippery wall, the bulk-boundary correspondence ensures the existence of topologically protected edge states \citep{PRB2011}.
When the wall is non-slip, we can still get a well-defined non-zero Chern number, but in this case, $k_{x}$ and $k_{y}$ need to be taken as complex numbers \citep{Non-Hermitian.Chern.Bands}.

The Rayleigh-Bénard convection is essentially a nonlinear system, so as a complement to the conclusions of this paper, it is necessary to continue to investigate the nonlinear effects on the topological properties of the bulk states.
We expect that weak nonlinear effects do not destroy the topological invariance of the system, partly because the topological properties are robust to perturbations and partly because previous numerical simulations imply this \citep{Favier_Knobloch.2020}. 
For a more detailed quantitative analysis, we need to evaluate the generalized geometric phase in the case of non-eigen states \citep{Nonlinear.Evolution.2003,Geometric.Phase.2005}.

\textit{Acknowledgement} This work has received financial support from the National Natural Science Foundation of China (NSFC) under grant NO. 92052102 and 12272006, and from the Laoshan Laboratory under grant NO. 2022QNLM010201.

\appendix
\section{Solutions satisfying realistic boundary conditions}\label{appA}

Unlike in Hermitian systems, the eigen frequencies and solutions of non-Hermitian systems are very sensitive to the boundary conditions. 
For simplicity, we assume that the horizontal direction is unbounded and the $z$-direction takes the no-slip boundary condition that
\begin{equation}
	u=v=w=\theta=0 \text{\quad on \quad} z=0,1.
\end{equation}

To ensure that our analyses above match a realistic situation, we construct solutions in the form that 
\begin{equation}
	\phi=e^{i(k_x x+k_y y-\omega t)}\sum_{i=1}^4 \lambda_i e^{ik_{zi} z}\left.\left|\psi_{i}\right.\right\rangle,
\end{equation}
where $\left.\left|\psi_{i}\right.\right\rangle$ is the eigenvector of Hamiltonian (\ref{H,Pr=1}) corresponding to $k_{zi}$.
According to our calculations above, when the system evolves along a closed path in its parameter space, $\left.\left|\psi_{1-4}\right.\right\rangle$ will all get the Berry phase of $2\pi\lambda$. As a result, the solution $\phi$ also gets the same Berry phase.
For a particular frequency $\omega(k_x,k_y)$, $k_{z1-4}$ come from the dispersion relation that ($Pr=1$)
\begin{equation}\label{dispersion_complex}
	\omega=\frac{\sqrt{k_{z}^{2}-\alpha(k_{x}^{2}+k_{y}^{2})}}{k}-iE k^2.
\end{equation}
Generally, the solved $k_{zi}$ is complex, and $k_{z3}=-k_{z1}$, $k_{z4}=-k_{z2}$. 

For the no-slip boundary condition, we have
\begin{equation}
	M\left[\begin{array}{c}
		\lambda_1\\
		\lambda_2\\
		\lambda_3\\
		\lambda_4
	\end{array}\right]=0,
\end{equation}
where
\begin{equation}\label{mat_M}
	M=\left[\begin{array}{cccc}
		1 & 1 & 1 & 1\\
		k_{z1} & k_{z2} & k_{z3} & k_{z4}\\
		e^{ik_{z1}} & e^{ik_{z2}} & e^{ik_{z3}} & e^{ik_{z4}}\\
		e^{ik_{z1}}k_{z1} & e^{ik_{z2}}k_{z2} & e^{ik_{z3}}k_{z3} & e^{ik_{z4}}k_{z4}
	\end{array}\right].
\end{equation}
To get a nontrivial solution of $\lambda_i(k_x,k_y)$, we have
\begin{equation}
	det(M)=0,
\end{equation}
and obtain that
\begin{equation}
	2 k_{z1} k_{z2} (-1 + \cos{k_{z1}} \cos{k_{z2}}) + (k_{z1}^2 + k_{z2}^2) \sin{k_{z1}} \sin{k_{z2}}=0.
\end{equation}
Obviously, $k_{z1}=k_{z2}$ satisfies the above equation, but this will give a trivial solution that $\lambda_i(k_x,k_y)=0$ and we do not choose it.
Combining with Eq.(\ref{dispersion_complex}), one can identify the nontrivial values of $k_{zi}$, $\omega(k_x,k_y)$ and $\lambda_i(k_x,k_y)$. As an example, for parameters $k_x=k_y=0.5$, $\alpha=1$ and $E=0.01$, numerical calculation gives that $\omega\approx0.9903-0.8108i$, $k_{z1}\approx8.9763-0.0195i$,  $k_{z2}\approx0.3998-0.6643i$ and
\begin{equation}
	\left[\begin{array}{c}
		\lambda_1\\
		\lambda_2\\
		\lambda_3\\
		\lambda_4
	\end{array}\right] =\mathcal{C} \left[\begin{array}{c}
		1\\
		0.9877-1.5056i\\
		0.9188-0.4422i\\
		-2.9066+1.9478i
	\end{array}\right],
\end{equation}
where $\mathcal{C}$ is an arbitrary constant.

\bibliographystyle{jfm}
\bibliography{mybib}

\end{document}